\documentclass{PoS}

\title{Recent results on two-photon physics at BABAR}

\ShortTitle{Recent results on two-photon physics at BABAR}

\author{\speaker{Vladimir DRUZHININ}%
\\
Budker Institute of Nuclear Physics\\
E-mail: \email{druzhinin@inp.nsk.su}}
			 
\abstract{
We present measurements of the $\gamma\gamma^\ast\to\pi^0$ transition form 
factor for the momentum transfer range $Q^2$=4--40 GeV$^2$ and the 
$\gamma\gamma^\ast\to\eta_c$ transition form factor for the range 
$Q^2$=2--50 GeV$^2$. The results of the measurement of the $\eta_c$ mass,  
total and two-photon widths are also presented.}

\FullConference{European Physical Society Europhysics Conference on High
Energy Physics\\
July 16-22, 2009\\
Krakow, Poland}

\begin{document}

\section{Introduction}
\begin{figure}
\includegraphics[width=.45\textwidth]{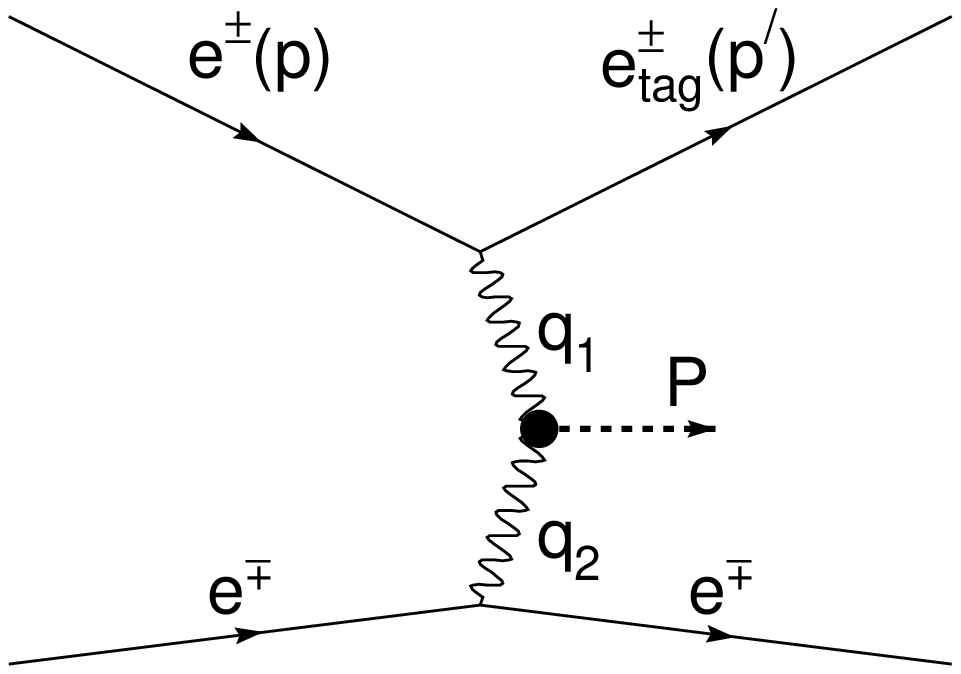}
\hfill
\includegraphics[width=.45\textwidth]{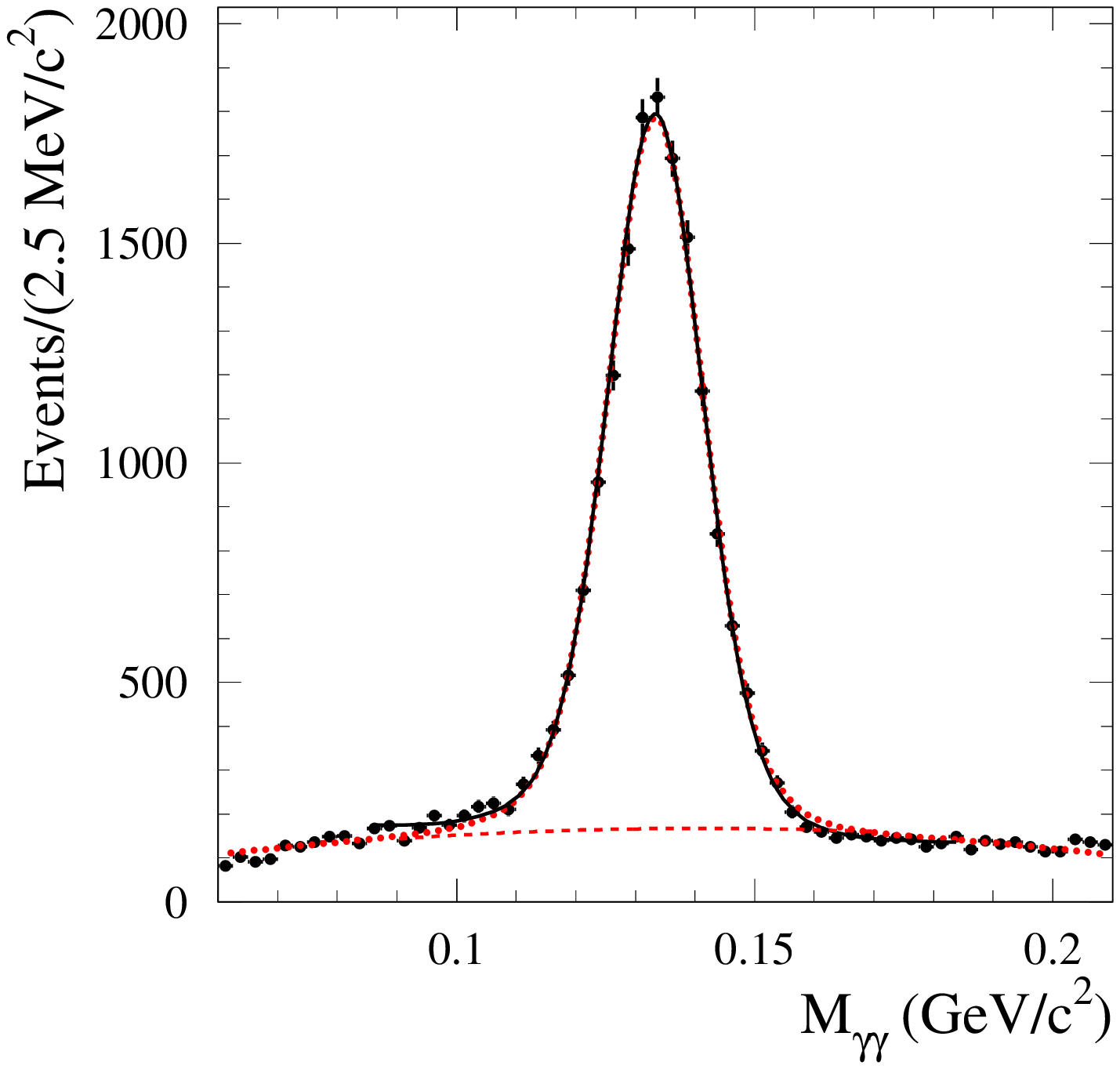}
\caption{{\bf Left panel:} The Feynman diagram for the process of the pseudoscalar meson
two-photon production.
{\bf Right panel:}
The two-photon invariant mass spectrum for data events with $4<Q^2<40$ GeV$^2$
and fitting curves.
\label{fig1}}
\end{figure}

The diagram for the process of the two-photon production of the pseudoscalar meson
is shown in Fig.~\ref{fig1}(left). The effect of strong interactions in this process
is described only one form factor $F(q_1^2, q_2^2)$ depending on the squared
momentum transfers to the electrons.

The electrons in such process are scattered predominantly at small angle. 
Therefore, the two-photon processes are usually studied in so called no-tag mode
with both final electrons undetected. In this case the virtual photons
are practically real, the momentum transfers squared are close to zero.
In no-tag mode the meson-photon transition form factor at zero $q^2$'s and the 
two-photon width of the meson are measured.
In single tag-mode the one of the final electron is detected.
The corresponding virtual photon is highly off-shell. From the measurement of
the cross section we extract more rich information: the dependence of
the meson form factor on $Q^2=-q_1^2$.

At large $Q^2$ perturbative QCD (pQCD) predicts that
the transition form factor can be represented as a convolution of a
calculable hard scattering amplitude for $\gamma\gamma^*\to q\bar{q}$
with a nonperturbative meson distribution amplitude (DA),
$\phi(x,Q^2)$~\cite{LB}. The latter can be interpreted as the amplitude for the 
transition of the meson with momentum $p$ into two quarks with momenta $px$ 
and $p(1-x)$. The experimental data on the transition form factor can be used
to test different phenomenological models for DA.

The cross section of the process $e^+e^-\to
e^+e^- P$ falls very rapidly with increase of $Q^2$ ($Q^{-6}$ for $\pi^0$).
Therefore, a precise measurement of the 
transition form factor can be performed only at high luminosity $e^+e^-$ 
machines. We present the results of the measurements of the transition
form factors for $\pi^0$ and $\eta_c$ mesons performed by the BABAR detector
at the PEP-II $e^+e^-$ collider. The results are based on data with integrated
luminosity of about 450 fb$^{-1}$ collected at the center-of-mass energy of 
10.6 GeV.
  The single-tag events are selected with detected and identified electron 
and with fully reconstructed $\pi^0$ or $\eta_c$. It is required that
the transverse momentum of electron-plus-meson system be low and
the missing mass in an event be close to zero.

\section{Measurement of the $\gamma^\ast\gamma\to\pi^0$ transition 
form factor~\cite{pi0ff}}
The $\pi^0$ meson is detected via its decay into two photons. The two-photon 
invariant mass spectrum for selected $\pi^0$ candidates is shown in 
Fig.~\ref{fig1}(right). The clear $\pi^0$ peak is seen. The main non-peaking 
background process is so called virtual Compton scattering (VCS), the precess 
$e^+e^-\to e^+e^-\gamma$ with one of the final electrons directed along the 
beam axis. 
The VCS photon together with a soft photon, for example from beam 
background, may give an invariant mass value close to the $\pi^0$ mass.
The peaking background comes from the process of two-photon production
of two $\pi^0$'s. This background is estimated from data and is about 10\% of
signal events. The total number of signal events determined from the fit to
the mass spectrum in Fig.~\ref{fig1}(right) is about 13000. This number is 
an order of magnitude large than the statistics of the previous measurement
of the form factor by CLEO~\cite{CLEO}. 

\begin{figure}
\includegraphics[width=.45\textwidth]{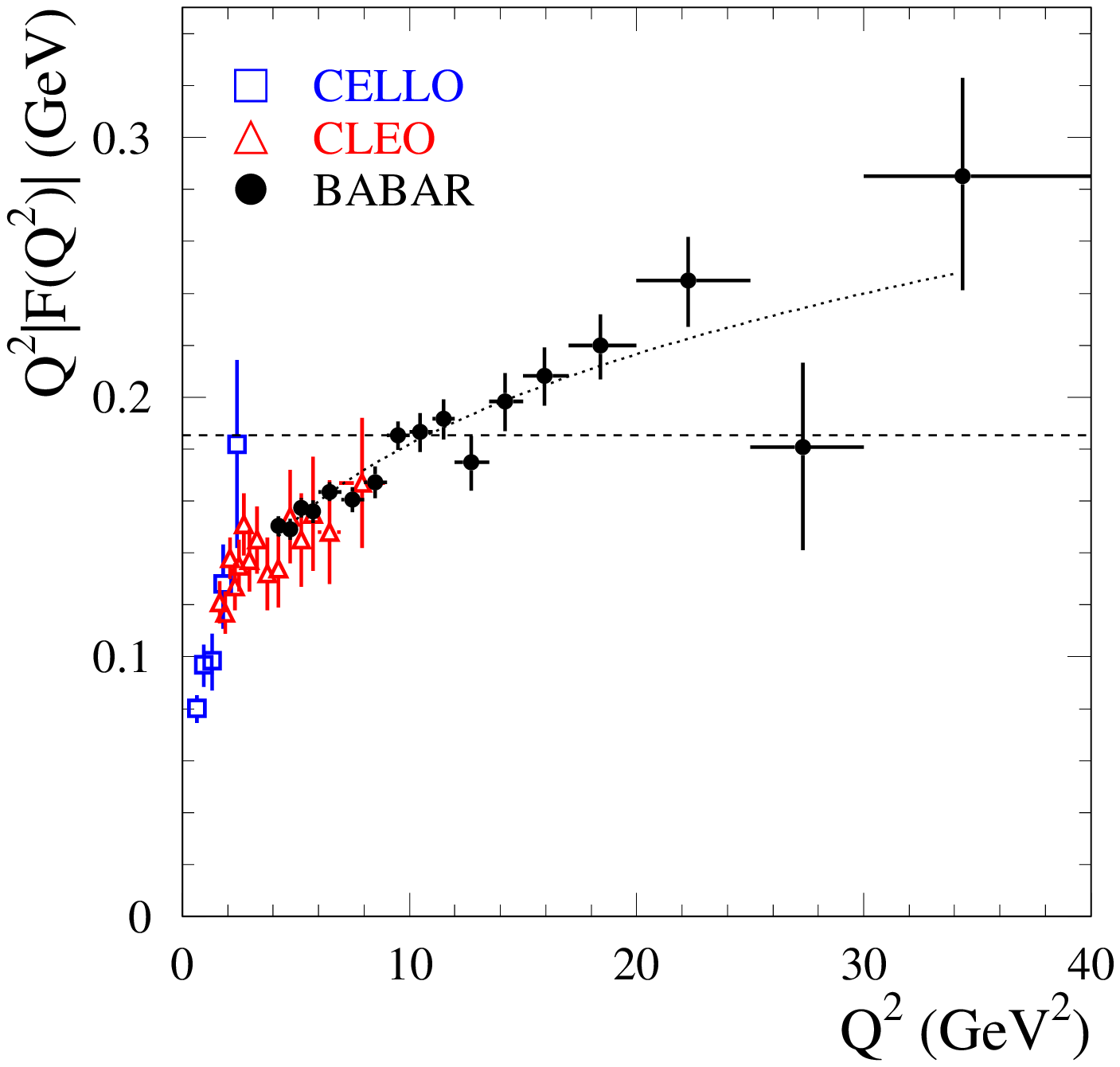}
\hfill
\includegraphics[width=.45\textwidth]{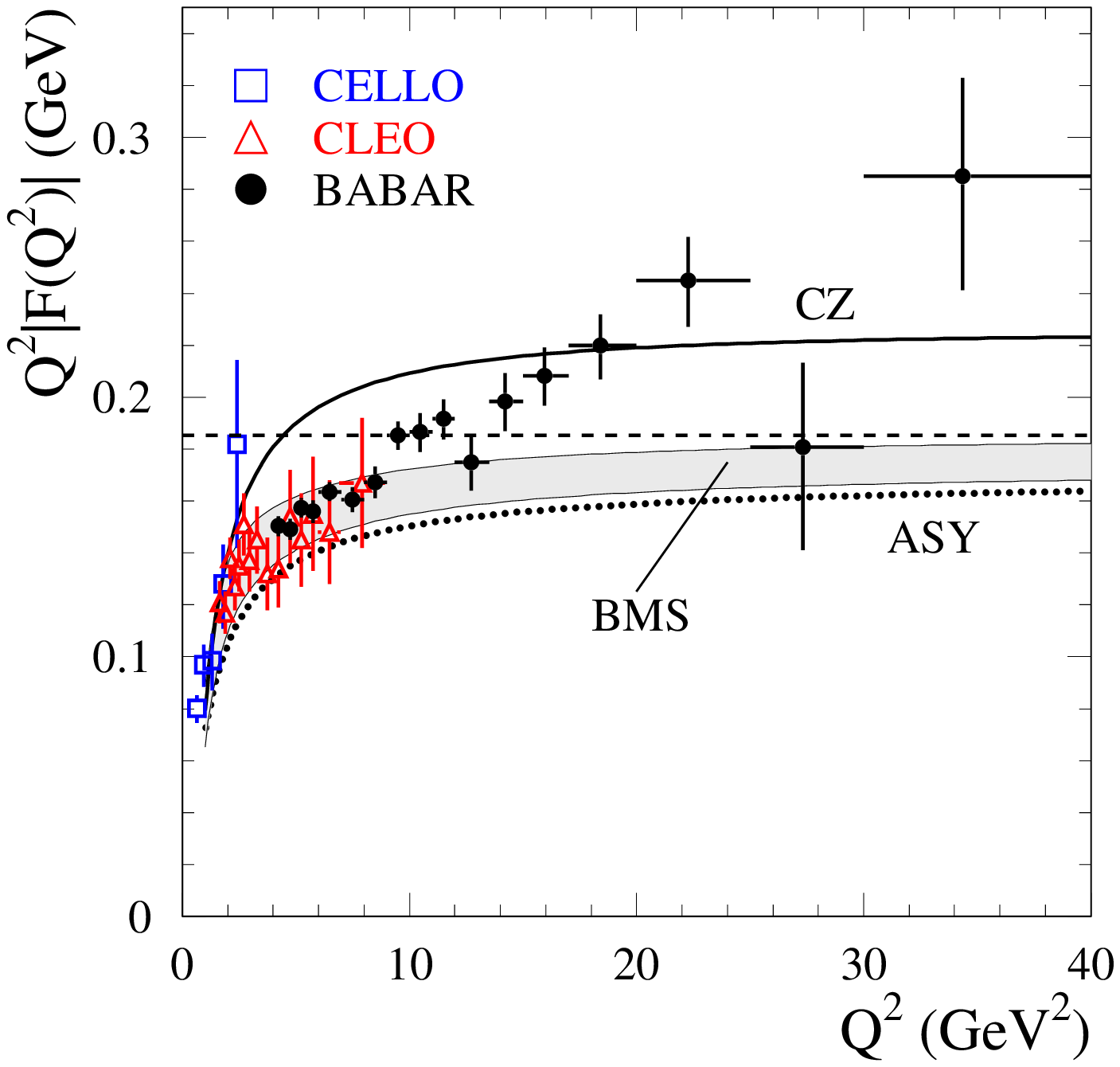}
\caption{
The $\gamma\gamma^\ast\to \pi^0$ transition form factor multiplied by
$Q^2$. The dashed line indicates the asymptotic limit for the form factor.
The dotted curve at the left panel shows the interpolation given by 
Eq.(2.1). 
The solid and dotted lines, and shaded band at the right panel show the 
predictions for the form factor for the CZ~\cite{CZ}, 
asymptotic (ASY)~\cite{ASY}, and BMS~\cite{BMS} models of the pion distribution 
amplitude, respectively. 
\label{fig3}}
\end{figure}

We measure the form factor in the $Q^2$ region from 4 to 40 GeV$^2$.
The lower $Q^2$ limit is determined by the detector acceptance for the
electron. For $Q^2 >40 $ GeV$^2$ we do not see evidence of a $\pi^0$ signal
over background. The data were divided into 17 $Q^2$ intervals.
For each $Q^2$ interval the mass spectrum is fitted by a sum of signal
and background distributions. From the measured $Q^2$ spectrum we determine
the differential cross section for $e^+e^-\to e^+e^-\pi^0$ and the
transition form factor. The result for the form factor is shown in
Fig.\ref{fig3}. The errors shown are combined statistical and $Q^2$-dependent
systematic. There is also $Q^2$-independent error equal to 2.3\%.
Main sources of the systematic uncertainties are background subtraction, 
data-MC simulation difference in the detector response, and the model 
uncertainty due to the unknown $q^2_2$ dependence of the form factor.

The comparison of our results  with previous measurements~\cite{CELLO,CLEO} 
is shown in Fig.~\ref{fig3}(left). In the $Q^2$ range from 4 to 9 GeV$^2$ our 
results are in reasonable agreement with the measurements by the CLEO 
collaboration~\cite{CLEO}, but have significantly better precision.

The horizontal dashed line indicates the asymptotic limit for the form factor.
The value of the asymptotic limit ($Q^2F(Q^2)=\sqrt{2}f_\pi\approx 0.185$ GeV)
is predicted by pQCD. The measured form factor exceeds the asymptotic
limit at $Q^2 > 10$ GeV$^2$. This is an unexpected behavior; most models for 
the pion DA give form factor approaching the limit from
below (see, e.g., Ref.~\cite{Stefanis} and references therein).
Our data in the range from 4 to 40 GeV$^2$ are well described by the 
function
\begin{equation}
Q^2|F(Q^2)|=A\left( \frac{Q^2}{10\mbox{ GeV}^2}\right)^\beta
\label{eqinter}
\end{equation} 
with $A=0.182\pm0.002$ GeV and $\beta=0.25\pm0.02$ (dotted line in
Fig.~\ref{fig3}(left)). The effective $Q^2$ dependence of the measured form factor
is $\sim 1/Q^{3/2}$.

Fig.~\ref{fig3}(right) demonstrates the comparison of our measurement with the 
result of the NLO QCD calculations performed by Bakulev, Mikhailov, and 
Stefanis~\cite{th6}
for the three models of the pion DA: 
asymptotic~\cite{ASY},
Chernyak-Zhitnitsky (CZ)~\cite{CZ}, and 
the DA derived from QCD sum rules with non-local condensates (BMS)~\cite{BMS}.
There is a large difference between data and theory in $Q^2$ dependence.
We conclude that all these models are inadequate for $Q^2 < 15$ GeV$^2$.
For $Q^2 > 20$ GeV$^2$ the theoretical uncertainties are expected to be
smaller. In this region our data lie above asymptotic limit and are
consistent with CZ model.  It should be noted that the CZ DA 
is widest of the three DA's discussed.

There are theoretical works which appeared after the publication of our 
result. Mikhailov and Stefanis~\cite{after0} argue that the growth of
form factor cannot be explained by higher-order pQCD and power
corrections. Other works~\cite{after1,after2,after3,after4} consider flat
or very wide pion DA. With such distribution amplitude the $Q^2$ 
dependence observed by BABAR is reproduced well.

\section{\boldmath Measurement of the $\gamma^\ast\gamma\to\eta_c$ transition form
factor}
\begin{figure}
\includegraphics[width=.45\textwidth]{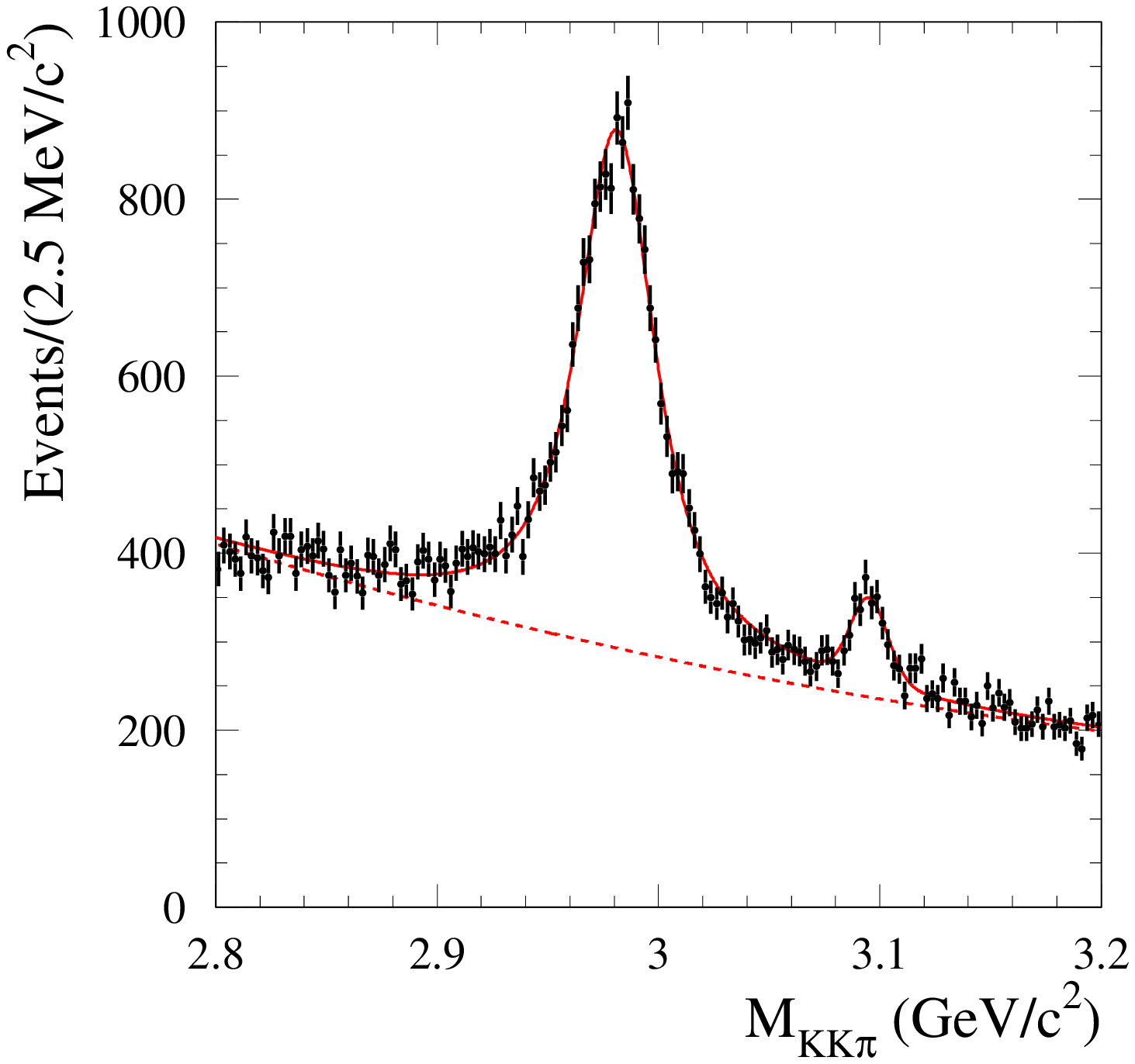}
\hfill
\includegraphics[width=.45\textwidth]{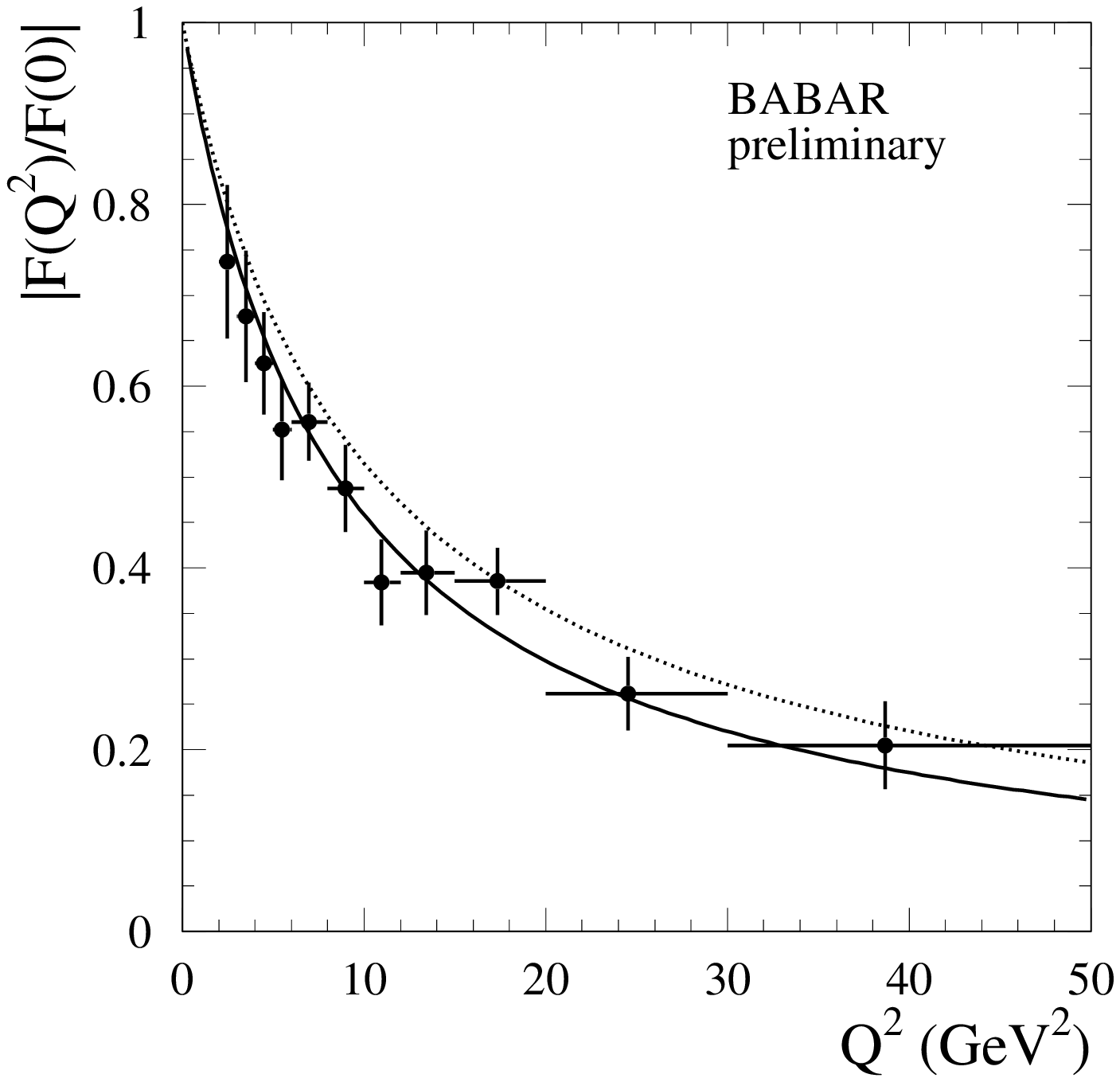}
\caption{{\bf Left panel:}
The $K_SK^\pm\pi^\mp$ invariant mass distribution and fitted
curve for no-tag data events.
{\bf Right panel:} The $\gamma\gamma^\ast\to \eta_c$ transition form factor
normalized to $F(0)$ (points with error bars). The solid curve
shows the interpolation given by a monopole form.
The dotted curve shows the leading order pQCD prediction from Ref.~\cite{th1}.
\label{fig4}}
\end{figure}

The two-photon $\eta_c$ production is studied both in no-tag and in single-tag
modes. The $\eta_c$ is reconstructed  via its decay to $K_SK^-\pi^+$.
The $KK\pi$ mass spectra for no-tag events is shown in Fig.~\ref{fig4}(left)
The $\eta_c$ and $J/\psi$ peaks are clearly seen.  The $J/\psi$'s are produced
in initial state radiation (ISR) process $e^+e^-\to J/\psi\gamma$.
From the fit to the mass spectrum we determine $\eta_c$ parameters:
$$m = 2982.2\pm0.4\pm1.5\mbox{ MeV/$c^2$},$$
$$\Gamma = 31.7\pm1.2\pm0.8 \mbox{ MeV},$$
$$\Gamma(\eta_c \to \gamma\gamma)B(\eta_c \to K\bar{K}\pi) =
0.379\pm0.009\pm0.030\mbox{ keV}.$$
These results are preliminary.
Main sources of systematic uncertainties on the mass and width are unknown 
background shape and possible interference between $\eta_c$ and non-resonant 
two-photon $KK\pi$ amplitudes. The uncertainty on the detection efficiency
dominates in the systematic  uncertainty of 
$\Gamma(\eta_c \to \gamma\gamma)B(\eta_c \to K\bar{K}\pi)$. 
The results for the mass and
width are in an agreement with the previous BABAR measurement~\cite{bb_etac}:
$m=2982.5\pm1.1\pm0.9$ MeV/$c^2$ and $\Gamma=34.3\pm2.3\pm0.9$
MeV, obtained using 88 fb$^{-1}$ data. The current PDG values for these
parameters are $m=2980.5\pm1.2$ MeV/$c^2$ and $\Gamma=27.4\pm2.9$ MeV~\cite{pdg}.
The obtained value of the product 
$\Gamma(\eta_c \to \gamma\gamma)B(\eta_c \to K\bar{K}\pi)$
agrees with the PDG value $0.44\pm0.05$ keV~\cite{pdg},
and also with the recent CLEO measurement $0.407\pm0.022\pm0.028$
keV~\cite{CLEO1}. 

We select $520\pm40\pm20$ single-tag $\eta_c$ events. This number can be compared
with $8\pm5$ events selected in the previous single-tag $\eta_c$ measurement
by L3~\cite{L3}. The single-tag data 
were divided into 11 $Q^2$ intervals. For each interval we fit to the $KK\pi$
mass spectrum and determine number of events with $\eta_c$. From the ratio
of the measured $Q^2$ spectrum to the number of the no-tag $\eta_c$ events we extract
the normalized $\eta_c$ transition form factor shown in Fig.~\ref{fig4}(right).
The errors shown are combined statistical and $Q^2$-dependent
systematic. There is also $Q^2$-independent error equal to 4.3\%.
Main source of the systematic error is the systematic uncertainty on detection
efficiency.

The form factor data are fitted by the monopole function
$|F(Q^2)/F(0)|=1/(1+Q^2/\Lambda)$.  The result of the fit is shown 
in Fig.~\ref{fig4}(right) by solid line. The pole parameter $\Lambda$ is found
to be $\Lambda=8.5\pm0.6\pm0.7\mbox{ GeV}^2$. This value is in reasonable
agreement with that expected from vector dominance, namely
$\Lambda=m^2_{J/\psi}=9.6$ GeV$^2$, and with the result of the lattice QCD 
calculation $\Lambda=8.4\pm0.4\mbox{ GeV}^2$~\cite{lqcd}.
The dotted curve in
Fig.~\ref{fig4}(right) shows results of the leading-order pQCD calculation of
Ref.~\cite{th1}. The data lie systematically below this prediction.
\section{Summary}
The $\gamma^\ast\gamma\pi^0$ transition form factor has been measured for $Q^2$
range from 4 to 40 GeV$^2$. The unexpected $Q^2$-dependence for the form factor 
is observed for $Q^2 > 10$ GeV$^2$. The data lie above the asymptotic limit. 
This indicates that the pion distribution amplitude should be wide.
The measurement stimulated development of new models for the form-factor
calculation, in particular, with flat distribution 
amplitude~\cite{after2,after3,after4}.

The $\gamma^\ast\gamma\eta_c$ transition form factor has been measured for $Q^2$
range from 2 to 50 GeV$^2$. The form factor data are well described by the monopole 
form with pole parameter about 9 GeV$^2$. The data are in reasonable agreement with
both QCD and VDM predictions.

\end{document}